\documentclass[12pt,preprint]{aastex}

\begin{document}
\def\ale{\mathrel{\hbox{\rlap{\hbox{\lower4pt\hbox{$\sim$}}}\hbox{$<$}}}}
\def\age{\mathrel{\hbox{\rlap{\hbox{\lower4pt\hbox{$\sim$}}}\hbox{$>$}}}}

\title{GRB~010921: Discovery of the First HETE Afterglow}

\author{
P.~A.~Price\altaffilmark{1,2},
S.~R.~Kulkarni\altaffilmark{1},
E.~~Berger\altaffilmark{1},
S.~G.~Djorgovski\altaffilmark{1},
D.~A.~Frail\altaffilmark{1,3},
A.~Mahabal\altaffilmark{1},
D.~W.~Fox\altaffilmark{1},
F.~A.~Harrison\altaffilmark{1},
J.~S.~Bloom\altaffilmark{1},
S.~A.~Yost\altaffilmark{1},
D.~E.~Reichart\altaffilmark{1},
A.~A.~Henden\altaffilmark{4},
G.~R.~Ricker\altaffilmark{5},
R.~van~der~Spek\altaffilmark{5},
K.~Hurley\altaffilmark{6},
J.-L.~Atteia\altaffilmark{7},
N.~Kawai\altaffilmark{8,9},
E.~Fenimore\altaffilmark{10} \&
C.~Graziani\altaffilmark{11}.
}

\altaffiltext{1}{Palomar Observatory, 105-24, California Institute of Technology, Pasadena, CA, 91125.}
\altaffiltext{2}{Research School of Astronomy \& Astrophysics, Mount Stromlo Observatory, via Cotter Road,
Weston, ACT, 2611, Australia.}
\altaffiltext{3}{National Radio Astronomy Observatory, P.O. Box O, Socorro, NM, 87801.}
\altaffiltext{4}{Universities Space Research Association / US Naval Observatory, Flagstaff Station,
P.O. Box 1149, Flagstaff, AZ, 86002-1149}
\altaffiltext{5}{Center for Space Research, Massachusetts Institute of Technology, Cambridge, MA, 02139-4307}
\altaffiltext{6}{University of California Space Sciences Laboratory, Berkeley, CA, 94720.}
\altaffiltext{7}{Centre d'Etude Spatiale des Rayonnements, CNRS/UPS, B.P. 4346, 31028 Toulouse Cedex 4,
France.}
\altaffiltext{8}{Department of Physics, Faculty of Science, Tokyo Institute of Technology, 2-12-1 Ookayama, Meguroku, Tokyo 152-8551, Japan.}
\altaffiltext{9}{RIKEN (The Institute of Physical and Chemical Research),
2-1 Hirosawa, Wako, Saitama 351-0198, Japan.} 
\altaffiltext{10}{MS D436, Los Alamos National Laboratory, Los Alamos, NM 87545.}
\altaffiltext{11}{Department of Astronomy and Astrophysics, University of Chicago, 5640 South Ellis Avenue,
Chicago, IL 60637.}

\begin{abstract}
We report the discovery of the optical and radio afterglow of
GRB~010921, the first gamma-ray burst afterglow to be found from a
localization by the High Energy Transient Explorer (HETE) satellite.
We present optical spectroscopy of the host galaxy which we find to be
a dusty and apparently
normal star-forming galaxy at $z = 0.451$.  The unusually steep
optical spectral slope of the afterglow can be explained by heavy
extinction, $A_V > 0.5$ mag,  along the line of sight to the GRB.  Dust
with similar $A_V$ for the 
the host galaxy as a whole appears to be required by the measurement
of a Balmer decrement in the spectrum of the host galaxy.  Thanks to
the low redshift, continued observations of the afterglow will enable
the strongest constraints, to date, on the existence of a possible
underlying supernova.  \end{abstract}



\section{Introduction}

The High Energy Transient Explorer (HETE-2) was successfully launched and
deployed on 2000 October 9 \citep{ricker+00}.  As the first satellite
entirely dedicated to the detection and study of gamma-ray bursts
(GRBs), HETE's primary mission is the localization of GRBs through their
prompt emission and the rapid relaying of the coordinates to ground-based
observers.  Here, we present the discovery of the afterglow of GRB~010921
\citep{ricker+01b},
the first afterglow to be identified through localization by HETE-2.


\section{Discovery of an Optical and Radio Transient}
\label{sec:discovery}

GRB~010921 triggered the FREGATE instrument on board HETE-2 on 2001
September 21.21934 UT as HETE trigger number 1761 \citep{ricker+01a}.
The GRB was detected by only one of HETE's two Wide-field X-ray
Monitors, and so localization was only possible in a single dimension.
Fortunately, the Interplanetary Network was able to provide
complementary localization and the resultant 310-square-arcmin error
box was speedily reported to the community via the GRB Coordinate
Network (GCN) Circulars \citep{hurley+01}.  We observed this error box
with the Large Format Camera (LFC) on the Hale 200-inch telescope at
Palomar Observatory, commencing 22 hours after the GRB, in the Sloan
Digital Sky Survey (SDSS; \citealt{yaa+00}) $r'$ filter.  Despite the
26-arcmin field-of-view (FOV) of the LFC, three pointings were required to
cover the entire error box.

A quick comparison of the images with the Digital Palomar Observatory
Sky Survey (\citealt{djorgovski+99}; also in prep.) failed to identify
an afterglow candidate within the large error box \citep{fox+01}.  Hoping to
identify the afterglow by its variability (i.e., decay), we re-observed
the error box with LFC five days later.

Reduction of data from LFC is complicated by the instrument's large FOV
which leads to substantial optical distortions in the focal plane, some
of which can be removed in software, but this does not correct for the
inevitable coma at the edges of the field.  Furthermore, this
problem is made worse
by our choice of on-chip binning (which however reduced the  file size
of the images).  We used the MSCRED package in the Image Reduction and
Analysis Facility (IRAF)\footnote{IRAF is distributed by the National
Optical Astronomy Observatories.} to bias-subtract, flat-field and
combine the individual images.

Comparison of the first- and second-epoch images was done through
PSF-matched image subtraction \citep{alard00}. 
Subtraction of images was rendered difficult due to the binning of
pixels. Nonetheless, we identified
source A ($\alpha_{2000}$ = 22$^{\rm h}$55$^{\rm m}$59$^{\rm s}$90,
$\delta_{2000}$ = +40$^{\circ}$55$^{'}$52$^{''}$9) and source B
($\alpha_{2000}$ = 22$^{\rm h}$56$^{\rm m}$15$^{\rm s}$09,
$\delta_{2000}$ = +41$^{\circ}$04$^{'}$48$^{''}$1), both of which were
within the error box and were clearly variable.  These 
positions are derived relative to USNO-A2.0 and are uncertain by
approximately 0.6 arcsec.

\begin{figure}[tbp]
\epsscale{0.3}
\plotone{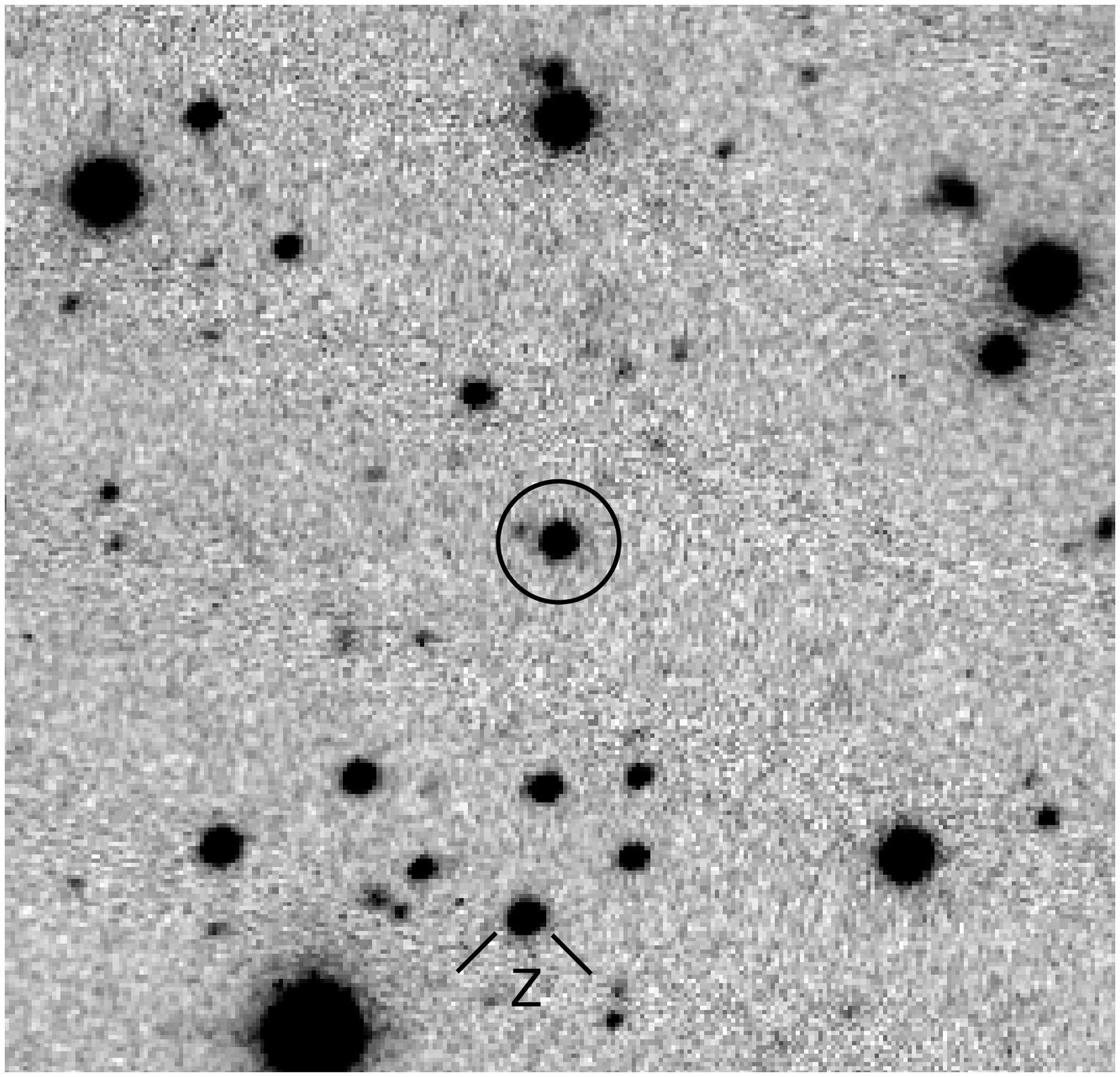}\hspace{0.1cm}\plotone{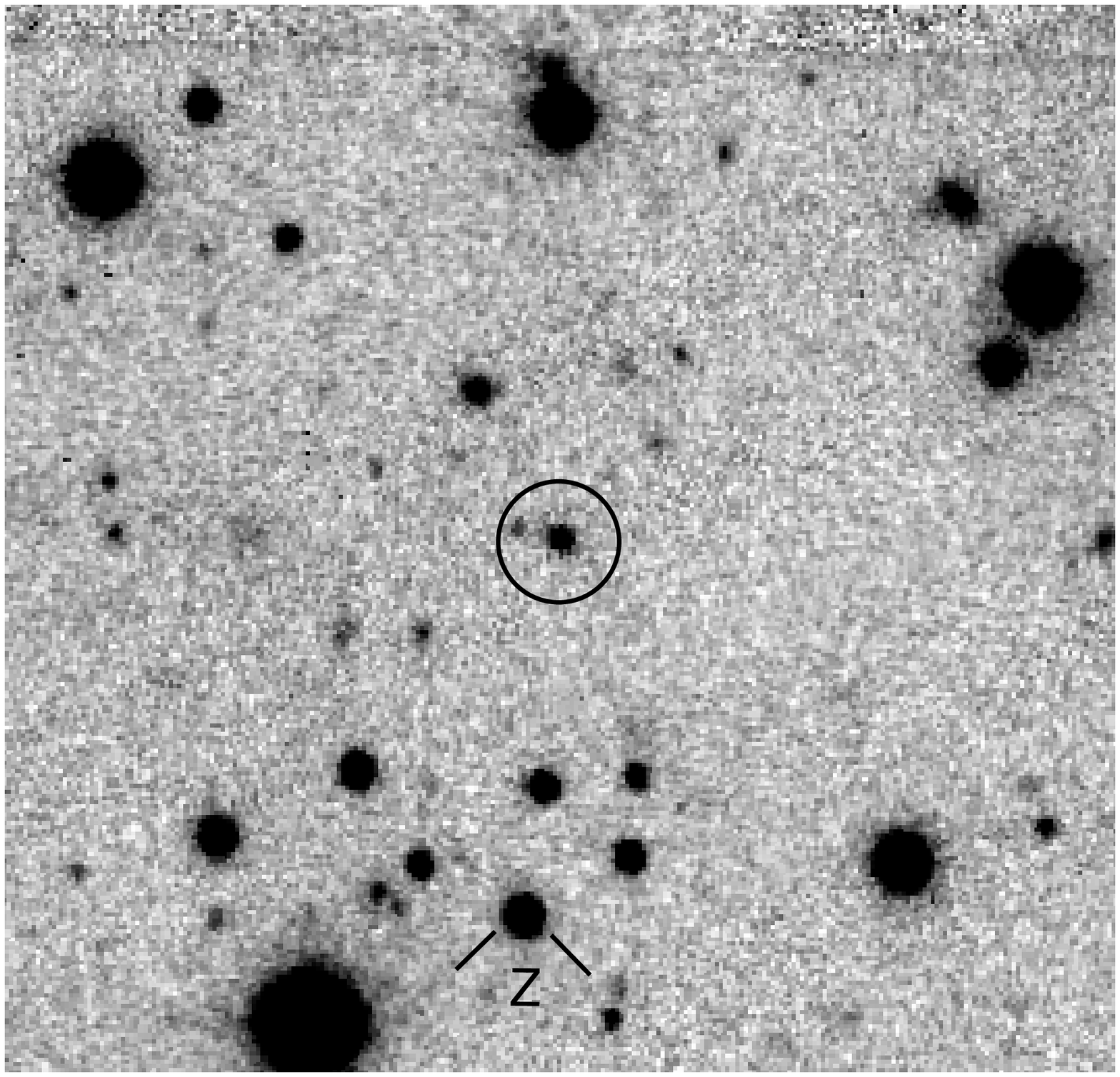}\hspace{0.1cm}\plotone{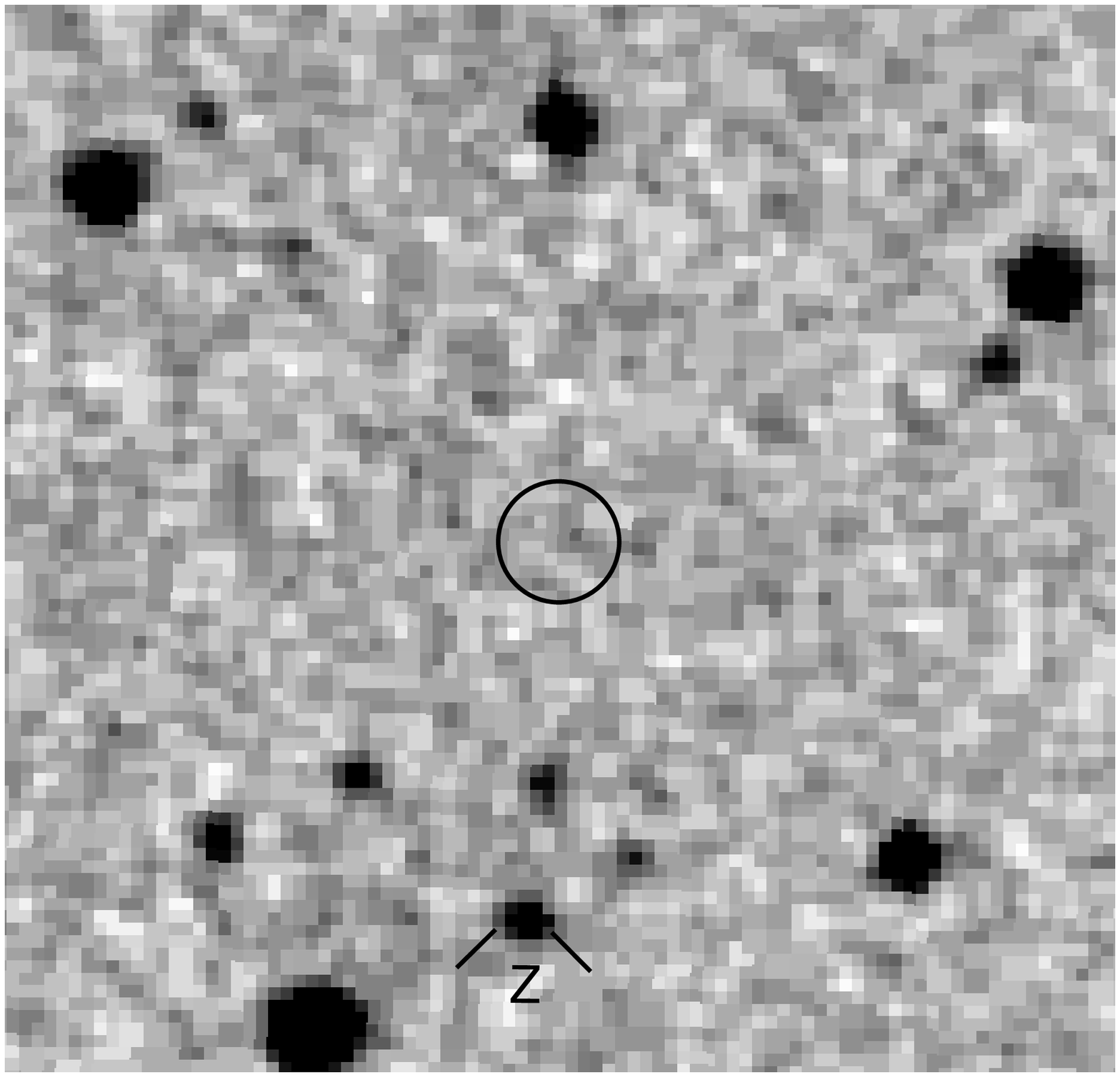}
\epsscale{1.0}
\caption{Identification of the optical afterglow of GRB~010921.  Left and
middle: $r'$ images from the Hale 200-inch with the LFC on 2001 September
22 and 27 respectively.  Right: Corresponding image from the DSS-2.  Each
image is 1.5-arcmin on a side, with North to the top and East to the left.
The position of the optical afterglow is circled.  Star Z (labeled), at
coordinates 
22$^{\rm h}$56$^{\rm m}$00$^{\rm s}$2 +40$^{\circ}$55$^{'}$22$^{''}$ (J2000)
is 2.65'' E and 30.35'' S of the afterglow, and for which we measure
$R = 21.169 \pm 0.043$ mag and $I = 19.537 \pm 0.025$ mag.}
\label{fig:FC}
\end{figure}

\begin{figure}[tbp]
\plotone{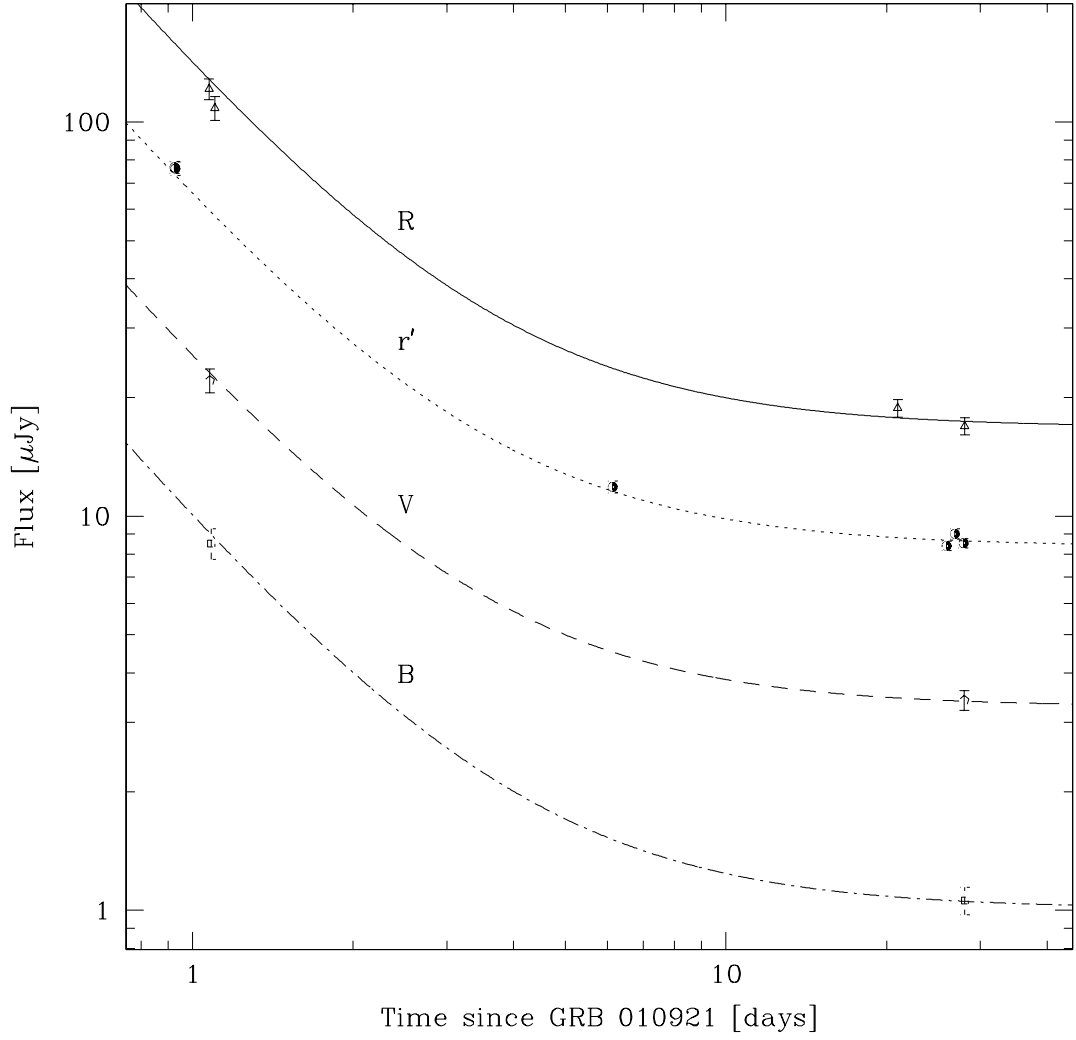}
\caption{The optical light curve of the afterglow of GRB~010921.  The
displayed fluxes have been corrected for Galactic extinction using
$E_{(B-V)} = 0.148$.  The lines indicated the best-fit power-law plus
galaxy model.}
\label{fig:lightcurve}
\end{figure}

In order to further investigate these two sources, we re-observed
the field on 2001 October 12th and 17th, 21 and 26 days after the GRB
respectively.  Photometry from these epochs revealed that source B had
increased in brightness, while source A had faded further and appeared
to be settling to a constant flux level; see Figures~\ref{fig:FC} and
\ref{fig:lightcurve}.
Source B, which is faintly visible at the plate limit of the second
Digitised Sky Survey\footnote{The Digitized Sky Surveys were produced at
the Space Telescope Science Institute under U.S. Government grant NAG W-2166.},
is likely a variable star or some other object unrelated to the GRB, while the
light curve of source A bears a similarity to the afterglows of other GRBs which
are observed to undergo a power-law decay followed by leveling off at the flux
level of the host galaxy.

Optical photometry was performed relative to stars in the field calibrated
through observations of numerous Landolt standards observed at the
USNO Flagstaff Station 1.0-meter telescope on 2001 September 22.
We estimate that these calibrations have a zero-point error of less
than 0.03 mag.  Reference star magnitudes in the SDSS photometric
system were derived through application of the transformations of
\citet{fukugita+96}.  The resulting photometry of source A can be found
in Table~\ref{tab:optical}.

A program of observations at the Very Large Array (VLA)\footnote{The National
Radio Astronomy Observatory (NRAO) is a facility of the National Science
Foundation operated under cooperative agreement by Associated Universities, Inc.
NRAO operates the VLA.}  were
begun commencing on 2001 October 17; see Table~\ref{tab:radio}.
All radio observations were obtained in the standard continuum mode
with $2\times 50$ MHz contiguous bands. We observed the extra-galactic
source J2255+420 for phase calibration, and 3C48 (J0137+331) and 3C286
(J1331+305) for flux calibration.  All data were reduced using the
Astronomical Image Processing System (AIPS).

Coincident with source A, we
identified a radio source at
$\alpha_{2000}=22^{\rm h}55^{\rm m}59^{\rm s}931 \pm 0.018$ and
$\delta_{2000}=
+40^{\circ}55^{'}52^{''}23 \pm 0.20$ and with 
a spectral slope of $\beta = 0.35 \pm
0.19$ (where $f_\nu \propto \nu^\beta$) between 4.86 and 22.5 GHz.
This value of $\beta$ is consistent with that expected
from the synchrotron emission of an afterglow ($\beta = 1/3$;
\citealt{spn98}), but does not exclude the possibility that the source
is a low-luminosity AGN.  Further observations indicate variability in
the source (Table~\ref{tab:radio}), which may be interpreted as
due to interstellar scintillation, a phenomenon commonly seen in radio
afterglows \citep{frail+97}.

\section{The Afterglow}
\label{sec:afterglow}

As noted above, the optical light curve, a decline followed by leveling
off, the variable radio emission and the flat radio spectrum, support
the identification of source A as the afterglow of GRB 010921. However,
the data are sparse and the possibility remains that source A could be
an AGN since AGNs also possess power-law spectra and can be strongly
variable. However, in section \ref{sec:spectroscopy} we note that the
optical spectrum of the host does not show any evidence for a central
engine. Furthermore, by comparing Palomar LFC images taken on 2001
September 22 (dominated by the optical transient) with those taken on
2001 October 17 (dominated by the galaxy), we find that the transient
is offset from the light centroid of the host galaxy by 0.351 $\pm$
0.049 arcsec, or 2.18 $\pm$ 0.30 kpc at the distance of the galaxy (see
\S \ref{sec:spectroscopy}).  Such angular and physical offsets with
respect to their host galaxy are typical of GRBs \citep{bkd02}.  Thus
from the totality of the radio and optical data we concluded that
source A is not an AGN but rather the afterglow of GRB~010921 and
announced it as such \citep{price+01b}.

After accounting for Galactic extinction 
towards this direction
($E_{(B-V)} = 0.148$; \citealt{sfd98}) we fit the 
optical photometric measurements to a power-law (afterglow) model,
$F_\nu \propto
t^{\alpha} \nu^{\beta}$. We obtain marginally-acceptable fits
($\chi^2$/DOF = 14.5/10), with $\alpha = -1.59 \pm 0.18$ and
$\beta = -2.22 \pm 0.23$.  \citet{pwh+02} report similar decay values
based on observations carried out with a number of telescopes.

The temporal decay slope is consistent with those seen in previous
optical afterglows, but the spectral slope is quite steep, suggestive
of extinction in the source frame.  Adopting standard afterglow theory
\citep{spn98}, we can deduce the intrinsic spectral slope from the
temporal data. However, this inference depends on the details
of the afterglow model (ambient gas, homogeneous versus 
inhomogeneous;
the electron distribution
power law index, $p$; the shape of the explosion, spherical versus
collimated outflow). The extinction along the line-of-sight to
the GRB, parametrized by $A_V$ in the rest frame of the source,
can then be obtained from the deviation of the observed
broad-band spectrum from that expected from the afterglow model.

The actual value of the derived $A_V$ is somewhat dependent on the
choice of both the extinction curve (Milky Way, LMC or SMC), and
the afterglow model (collimated or isotropic ejecta; cooling break,
redward or blueward of the optical band over the course of the
observations) but, as can be seen from Table~\ref{tab:extinction} it
is clear that $A_V$ is in excess of 0.5 mag. The best fit model has
a spherical outflow with the cooling frequency below the optical band
by day 1 and an LMC-like extinction curve ($\chi^2$/DOF = 15.5/11).

\section{The Host Galaxy}
\label{sec:spectroscopy}

\begin{figure}[tbp]
\epsscale{0.5}
\centering 
\leavevmode 
\includegraphics[angle=-90,scale=0.6]{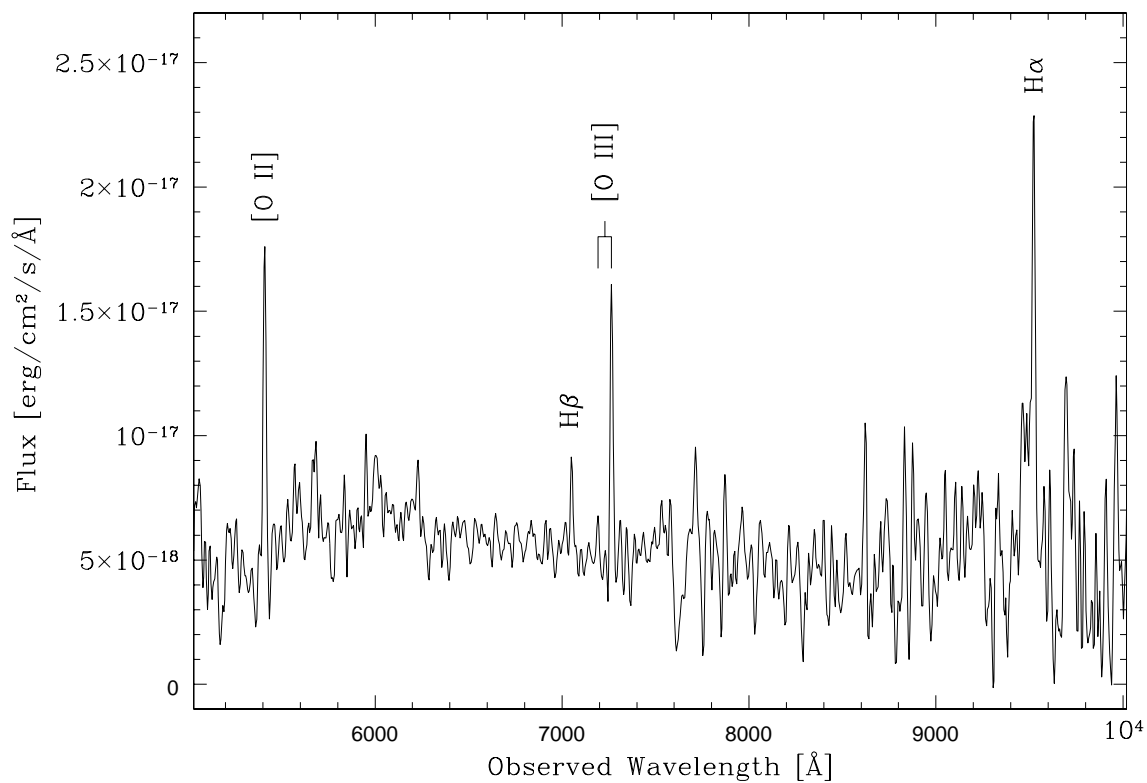}
\caption{The spectrum of the host galaxy of GRB 010921.  The spectrum has
been smoothed with a Gaussian of FWHM = 12 \AA, approximately the
instrumental resolution. The continuum longward of 8500\AA\ suffers
from poor sky subtraction.
Emission lines corresponding to [O II], H$\beta$,
[O III] and H$\alpha$ (labeled) are clearly detected, corresponding to a
mean redshift of $z = 0.4509$.}
\label{fig:spectrum}
\end{figure}

Using the Double Spectrograph \citep{og82} at the Cassegrain focus of the
Hale 200-inch telescope we obtained a spectrum of the host galaxy
(Figure~\ref{fig:spectrum}).  We used a
1-arcsec wide long-slit, at a position angle
$132^\circ$, close to the parallactic angle,
and a 158 lines mm$^{-1}$ grating, giving an effective instrumental
resolution FWHM $\approx 12 $\AA, and an approximate wavelength coverage
5,000--10,000~\AA\ on the red side.  We obtained 2 exposures of 1800-s
each, with a small dither along the slit.  Exposures of arc lamps
were obtained for calibrations and small wavelength shifts due to the
instrument flexure were removed by fitting the wavelengths of night
sky lines.  The net resulting wavelength scatter is $\sim 0.4$ \AA.
Flux calibration was made through observations of the spectrophotometric
standard BD$+17^\circ 4708$ from Oke \& Gunn (1983)\nocite{og83}.

Several emission lines are detected (see Table~\ref{tab:lines}),
corresponding to [O II]$\lambda$3727, H$\beta$, [O
III]$\lambda\lambda$4959,5007 and H$\alpha$, at a mean redshift of
$z = 0.4509 \pm 0.0004$.  We note that the widths of the lines,
ranging from 10 \AA\ to 13 \AA, are consistent with the instrumental
resolution and thus there is  no reason to invoke an AGN component to
this host galaxy.
Furthermore, the line flux ratios $F_{5007}/F_{4861}$ and
$F_{3727}/F_{5007}$ place the host galaxy squarely in the range
corresponding to normal HII galaxies on the identification plot of
\citet{bpt81}.

In estimating the properties of the host galaxy, we adopt a standard
flat cosmology with $H_0 = 65$ km s$^{-1}$ Mpc$^{-1}$, $\Omega_{\rm M}
= 0.3$ and $\Omega_\Lambda = 0.7$.  From the observed Balmer decrement
and an LMC extinction curve, we calculate an optical depth at H$\beta$
due to dust of 1.51 (assuming an intrinsic H$\alpha$ to H$\beta$ flux
ratio of 2.87; \citealt{mathis83}); note, we have not accounted for
Balmer absorption lines in the host galaxy.   The corresponding $A_V$
which applies to the galaxy as a whole is 1.3 mag. This value of
$A_V$ is comparable to the extinction along the line-of-sight to the
GRB afterglow and thus most likely the extinction observed in the
afterglow is due to interstellar (as opposed to circumburst) dust.

There are two other GRB host galaxies for which Balmer line fluxes have
been published, the hosts of GRBs 980703 \citep{djorgovski+98} and
990712 \citep{vreeswijk+01}, for which we calculate optical depths at
H$\beta$ of approximately 0.1 and 2.41, respectively, using the above
method.  We note that the Balmer decrement observed for the host of
GRB~010921 falls squarely within the range observed for these other two
GRB hosts.  

We now infer the  metallicity (oxygen abundance) using the empirical
relation between oxygen abundance and $\log R_{23} \equiv \log\left(([O
II] + [O III])/H\beta\right)$ \citep{kkp99}. We find $\log R_{23}= 1.05 \pm
0.18$ which is close to  the knee of this relation and thus yields an
unambiguous measure of the oxygen abundance, 12 + log (O/H) of $8.35
\pm 0.25$.  The rest-frame equivalent width of the [O II] line, $34.5
\pm 3.5$ \AA, is typical for field galaxies \citep{hogg+98}, which have
equivalent widths spanning 0 -- 50 \AA.  Furthermore, using the [O II]
and H$\alpha$ line fluxes and the empirical star-formation rate (SFR)
estimators of \citet{kennicutt98} we derive extinction-corrected SFR
of 8~M$_{\odot}$ yr$^{-1}$ and 6~M$_{\odot}$ yr$^{-1}$ respectively.
All in all, we conclude that GRB 010921 occured in a normal, star-forming
galaxy with an oxygen abundance approximately one-quarter solar.

We also observed the host galaxy with the Near InfraRed Camera (NIRC;
\citealt{ms94}) on the Keck I telescope on 2001 November 1 UT in the
$JHK_s$ bands under photometric conditions.  The individual frames were
flat-fielded and then sky-subtracted and combined using the DIMSUM
package in IRAF \citep{eisenhardt+99}.  Flux calibration was performed
through observation of \citet{persson+98} standards SJ9126 and SJ9134.
We estimate that the calibration is accurate to within 0.05 mag.  The
resulting magnitudes of the host galaxy, measured within a 1.5-arcsec
aperture, are also displayed in table~\ref{tab:optical}.  From the flat
spectrum in the NIR, we estimate that the stellar population of the
host is less than 1 Gyr in age.

From the FREGATE 8--400 keV fluence of $1.5 \times 10^{-5}$ erg/cm$^2$
\citep{ricker+01b}, we derive the isotropic-equivalent prompt energy
release (without any k correction) of 
$9.0 \times 10^{51}$ erg.   However, it is now generally
appreciated that GRBs are beamed or collimated \citep{frail+01} and thus
the isotropic-equivalent value is an upper limit. Unfortunately, the
paucity of the afterglow data do not place significant constraint on
the opening angle of the explosion (see Table~\ref{tab:extinction}).

\section{Conclusion}
\label{sec:conclusion}

Here we report the first afterglow resulting from an HETE localization
of a GRB.  The afterglow in itself is similar to afterglows studied to
date. However, the GRB occured in a low-redshift, $z=0.451$, dusty
star-forming galaxy. Such low redshift GRBs are valuable for two
reasons. First and foremost, such low-redshift events offer the best
opportunity to study the explosion physics including the opportunity to
constrain the presence of (or detect) any underlying supernova.
Second, the low redshift allowed us to determine the metallicity of the
host galaxy with relative ease.

It is perhaps significant to note that the afterglow was discovered
through the use of image differencing.  We suggest that the use of this
technique may be a more robust method of identifying GRB afterglows
than the traditional manual comparison with sky survey images, since it
enables the detection of an afterglow superimposed on a bright host
galaxy. This is particularly important for low redshift GRBs.  We
wonder whether the traditional approach --- looking for an isolated
transient --- could be the cause of failure to identify optical
afterglows in some GRBs.

\acknowledgements

We thank Martha Haynes for the generous donation of her Hale 200-inch
time without which the discovery of this afterglow would not have been
possible.  We thank Rick Burrus for his help with 200-inch
observations.  PAP gratefully acknowledges an Alex Rodgers Travelling
Scholarship.  DAF thanks Caltech for the hospitality shown by Caltech
during his sabbatical leave. 
SRK and SGD thank NSF for support of our ground-based GRB
observing program.  JSB acknowledges support from the Hertz Foundation
in the form of a fellowship.  KH is grateful for {\it Ulysses} support
under Contract JPL 958059, and for HETE support under Contract
MIT-SC-R-293291.

\clearpage

\begin{deluxetable}{ccccc}
\footnotesize
\tablecolumns{5}
\tablewidth{0pt}
\tablecaption{\label{tab:optical}Optical observations of the afterglow of
GRB~010921.}
\tablehead{\colhead{Date (UT)} & \colhead{Filter} & \colhead{Magnitude} & \colhead{Telescope} & \colhead{Lead Observer}}
\startdata
2001 Sep 22.144   &	$r'$  &	19.593   $\pm$	0.033 &	P200 + LFC	& Haynes/Burrus	\\
2001 Sep 22.148   &	$r'$  &	19.601   $\pm$	0.039 &	P200 + LFC	& Haynes/Burrus	\\
2001 Sep 22.289   &	$I$   &	18.854   $\pm$	0.064 &	USNOFS1.0	& Henden	\\
2001 Sep 22.293   &	$R$   &	19.669   $\pm$	0.060 &	USNOFS1.0	& Henden	\\
2001 Sep 22.298   &	$V$   &	20.296   $\pm$	0.070 &	USNOFS1.0	& Henden	\\
2001 Sep 22.304   &	$B$   &	21.159   $\pm$	0.088 &	USNOFS1.0	& Henden	\\
2001 Sep 22.313   &	$U$   &	20.910   $\pm$	0.205 &	USNOFS1.0	& Henden	\\
2001 Sep 22.321   &	$R$   &	19.790   $\pm$	0.069 &	USNOFS1.0	& Henden	\\
2001 Sep 27.354   &	$r'$  &	21.620   $\pm$	0.036 &	P200 + LFC	& Djorgovski	\\
2001 Oct 12.226   &	$R$   &	21.693   $\pm$	0.053 &	P200 + LFC	& Bloom	\\
2001 Oct 17.145   &	$r'$  &	21.994   $\pm$	0.028 &	P200 + LFC	& Mahabal	\\
2001 Oct 17.165   &	$i'$  &	21.660   $\pm$	0.022 &	P200 + LFC	& Mahabal	\\
2001 Oct 18.088   &	$r'$  &	21.918   $\pm$	0.030 &	P200 + LFC	& Mahabal	\\
2001 Oct 18.110   &	$i'$  &	21.670   $\pm$	0.021 &	P200 + LFC	& Mahabal	\\
2001 Oct 19.178   &	$g'$  &	22.947   $\pm$	0.043 &	P200 + LFC	& Mahabal	\\
2001 Oct 19.109   &	$r'$  &	21.977   $\pm$	0.028 &	P200 + LFC	& Mahabal	\\
2001 Oct 19.130   &	$i'$  &	21.650   $\pm$	0.030 &	P200 + LFC	& Mahabal	\\
2001 Oct 19.149   &	$z'$  &	21.419   $\pm$	0.043 &	P200 + LFC	& Mahabal	\\
2001 Oct 19.253   &	$B$   &	23.423   $\pm$	0.079 & P60		& Fox	\\
2001 Oct 19.206   &	$V$   &	22.324   $\pm$	0.057 & P60		& Fox	\\
2001 Oct 19.272   &	$R$   &	21.807   $\pm$	0.051 & P60		& Fox	\\
2001 Nov 01.287   &	$K_s$ & 19.069   $\pm$	0.037 & KI + NIRC	& Kulkarni	\\
2001 Nov 01.345   &	$H$   & 19.750   $\pm$	0.035 & KI + NIRC	& Kulkarni	\\
2001 Nov 01.316   &	$J$   & 20.338   $\pm$	0.019 & KI + NIRC	& Kulkarni	\\
\enddata
\tablecomments{These measurements have not been corrected for Galactic
extinction.  The error in the measurement includes both statistical and
systematic errors.  A 2.5-arcsec aperture was used for optical measurements,
and a 1.5-arcsec aperture for the NIR measurements.  Telescopes are: P200 --- Hale
Palomar 200-inch; USNOFS1.0 --- USNO Flagstaff Station 1.0-metre; P60 --- Palomar
60-inch; KI --- Keck I.}
\end{deluxetable}

\clearpage

\begin{deluxetable}{lcc}
\tabcolsep0in
\footnotesize
\tablewidth{0pt}
\tablecaption{Radio Observations of GRB~010921 made with the Very Large Array.\label{tab:radio}}
\tablehead {
\colhead {Epoch}      &
\colhead {$\nu_0$} &
\colhead {S$\pm\sigma$} \\
\colhead {(UT)}      &
\colhead {(GHz)} &
\colhead {($\mu$Jy)}
}
\startdata
2001 Oct 17.15 & 4.86 & 188 $\pm$ 25	\\
2001 Oct 17.15 & 8.46 & 222 $\pm$ 16	\\
2001 Oct 17.15 & 22.5 & 330 $\pm$ 90	\\
2001 Oct 18.23 & 8.46 & 229 $\pm$ 22	\\
2001 Oct 19.02 & 4.86 & 100 $\pm$ 28	\\
2001 Oct 24.02 & 22.5 & 37  $\pm$ 95	\\
2001 Oct 26.04 & 4.86 & 91  $\pm$ 36	\\
2001 Oct 28.99 & 4.86 & 140 $\pm$ 28	\\
2001 Oct 28.99 & 8.46 & 158 $\pm$ 26	\\
2001 Oct 29.99 & 8.46 & 123 $\pm$ 26	\\
\enddata
\tablecomments{The columns are (left to right), UT date of the
start of each observation, observing frequency, and peak flux density
at the best fit position of the radio transient, with the error given as
the root mean square noise on the image.}
\end{deluxetable}

\clearpage

\begin{deluxetable}{cccccc}
\footnotesize
\tablecolumns{6}
\tablewidth{0pt}
\tablecaption{\label{tab:lines}Lines identified in the spectrum of
the host galaxy of GRB~010921.}
\tablehead{\colhead{ $\lambda_{\rm obs}$ (\AA)} & \colhead{Line} & \colhead{$F_{\rm obs}$} & \colhead{EW (\AA)} & \colhead{GW (\AA)} & \colhead{$F_{\rm corr}$}}
\startdata
5408.77$\pm$0.57   &	[O II]   & 2.42$\pm$0.23 & 50.1$\pm$5.1 & 11.5$\pm$1.1 & 24.4$\pm$2.3	\\
5613.6		   &	[Ne III] & $< 0.98 $	 & $< 13.8$	& \ldots       & $< 8.9$	\\
7051.35$\pm$1.00   &	H$\beta$ & 0.69$\pm$0.11 & 13.2$\pm$2.3 & 10.2$\pm$1.9 & 3.12$\pm$0.50	\\
7192.81$\pm$1.76   &	[O III]  & 0.41$\pm$0.13 & 9.2$\pm$3.0  & 10.8$\pm$3.3 & 1.75$\pm$0.55	\\
7264.48$\pm$0.55   &	[O III]  & 2.18$\pm$0.14 & 43.5$\pm$3.5 & 12.9$\pm$1.1 & 9.11$\pm$0.59	\\
9522.46$\pm$1.22   &	H$\alpha$& 3.29$\pm$0.59 & 52$\pm$13    & 12.3$\pm$3.7 & 8.9$\pm$1.6	\\
\enddata
\tablecomments{Left to right, the columns are the observed wavelength of the
line, line identification, observed flux corrected for Galactic extinction
using $E_{(B-V)} = 0.148$, observed equivalent width, observed Gaussian width
and flux corrected for extinction in the host galaxy using an optical depth
at H$\beta$ of 1.51 and an LMC extinction curve.  Fluxes are in units of
$10^{-16}$ erg/cm$^2$/s. Given the suggestion that GRB host galaxies
may exhibit strong [Ne III] emission \citep{bdk01} we have included
an entry for this line.}
\end{deluxetable}

\clearpage

\begin{deluxetable}{cccccccc}
\footnotesize
\tablecolumns{6}
\tablewidth{0pt}
\tablecaption{\label{tab:extinction}Results of extinction curve fits to
the optical afterglow photometry. }
\tablehead{\colhead{Afterglow} & \colhead{Collimated?} & \multicolumn{2}{c}{MW ($c_2 = 2/3$)} & \multicolumn{2}{c}{LMC ($c_2 = 4/3$)} & \multicolumn{2}{c}{SMC ($c_2 = 7/3$)} \\
\colhead{Model} & \colhead{yes/no} & \colhead{$A_V$ (mag)} & \colhead{$\chi^2$} & \colhead{$A_V$ (mag)} & \colhead{$\chi^2$} & \colhead{$A_V$ (mag)} & \colhead{$\chi^2$} }
\startdata
$\nu < \nu_c$, $p = 3.1$ & no & 1.05 & 18.8 & 0.73 & 16.1 & 0.95 & 16.1 \\
$\nu > \nu_c$, $p = 2.8$ & no & 0.76 & 16.8 & 0.52 & 15.5 & 0.66 & 15.5 \\
$\nu > \nu_c$, $p = 1.6$ & yes& 1.29 & 21.0 & 0.89 & 16.7 & 1.17 & 16.8 \\
$\nu < \nu_c$, $p = 1.6$ & yes& 1.75 & 26.0 & 1.20 & 17.7 & 1.60 & 18.0 \\
\enddata
\tablecomments{The extinction curve formulation of Reichart (1999) was used
in performing these fits.  The models allow for both collimated ($t_{jet} < 1$
d) and isotropic ($t_{jet} > 25$ d) ejecta, and for the cooling break
to be either redward ($\nu > \nu_c$) or blueward ($\nu < \nu_c$) of the
optical bands over the period of observations of the afterglow (1 d $< t <$
25 d).  $A_V$ is source frame V-band extinction; typical
errors are 0.15 mag (MW), 0.10 mag (LMC) and 0.20 mag (SMC), not including
co-variance.  Each fit had 11 degrees of freedom.}
\end{deluxetable}

\clearpage

\clearpage

\clearpage

\clearpage

\end{document}